# Physics of nanocoulomb-class electron beams in laser-plasma wakefields

J. Götzfried,[1,2] A. Döpp,[1,2,*] M. F. Gilljohann,[1,2] M. Foerster,[1] H. Ding,[1,2]
S. Schindler,[1,2] G. Schilling,[1] A. Buck,[2] L. Veisz,[2,3] and S. A. Karsch[1,2,†]

[1]*Ludwig-Maximilians-Universität München, Am Coulombwall 1, 85748 Garching, Germany*
[2]*Max-Planck-Institut für Quantenoptik, Hans-Kopfermann-Str. 1, 85748 Garching, Germany*
[3]*Department of Physics, Umeå University, SE-901 87, Umeå, Sweden*

Laser wakefield acceleration (LWFA) and its particle-driven counterpart, plasma wakefield acceleration (PWFA), are commonly treated as separate, though related branches of high-gradient plasma-based acceleration. However, novel proposed schemes are increasingly residing at the interface of both concepts where the understanding of their interplay becomes crucial. Here, we report on experiments covering a wide range of parameters by using nanocoulomb-class quasi-monoenergetic electron beams from LWFA with a 100-TW-class laser. Based on a controlled electron injection, these beams reach record-level performance in terms of laser-to-beam energy transfer efficiency (up to 10 %), spectral charge density (regularly exceeding $10\,\mathrm{pC\,MeV^{-1}}$) and divergence (1 mrad full width at half maximum divergence). The impact of charge fluctuations on the energy spectra of electron bunches is assessed for different laser parameters, including a few-cycle laser, followed by a presentation of results on beam loading in LWFA with two electron bunches. This scenario is particularly promising to provide high-quality electron beams by using one of the bunches to either tailor the laser wakefield via beam loading or to drive its own, beam-dominated wakefield. We present experimental evidence for the latter, showing a varying acceleration of a low-energy witness beam with respect to the charge of a high-energy drive beam in a spatially separate gas target. With the increasing availability of petawatt-class lasers the access to this new regime of laser-plasma wakefield acceleration will be further facilitated, thus providing new paths towards low-emittance beam generation for future plasma-based colliders or light sources.

## I. INTRODUCTION

Plasma-based high-gradient wakefield accelerators have attracted significant interest in recent years due to their potential for a significant reduction in size and cost of future accelerators[1]. They seem particularly attractive as drivers for compact high-brightness photon sources[2,3], but may also play a role as building blocks for TeV-scale high-energy physics machines[4]. Important milestones include the generation of mono-energetic electron beams[5–7], sustained acceleration of more than a GeV[8–13], as well as generation and application of spatially-coherent, ultra-short X-ray sources based on the accelerated particle beams[14–19]. So far, this novel generation of accelerators has been used to drive applications such as X-ray imaging[20–23] and tomography[24–26], as well as femtosecond electron diffraction[27], femtochemistry[28] and ultrafast spectroscopy[29,30].

The technology relies on an intense drive beam, either a laser pulse or a particle bunch, ploughing through a plasma medium and pushing the electrons aside by their ponderomotive force or Coulomb repulsion, respectively. This sets up charge separation fields that pull electrons back and cause them to oscillate around their equilibrium position, but as the driver travels though the medium with a velocity close to the speed of light, the field structure ("wakefield") follows it at the same speed. Electrons injected into this moving wakefield can be accelerated as a so-called witness bunch. The accelerating field structure has a typical scale given by the plasma wavelength ($\lambda_\mathrm{p}$), and both accelerating and focusing gradients are several orders of magnitude larger than in radiofrequency (RF) accelerators, leading to very dense and ultra short particle bunches (some tens of femtoseconds duration). Depending on the driver type, the process is either called laser wakefield acceleration (LWFA)[31] or plasma wakefield acceleration (PWFA)[32]. While the former can be studied at many high-intensity laser laboratories at drive pulse powers starting at the TW level, the latter typically requires drive bunches that can only be provided by a national-laboratory scale RF accelerator. Despite the fact that both types differ in certain details of wakefield excitation, ionization and propagation[33], they begin to merge in the case of high-power LWFA. The beam loading effect[34], which describes the dependence of the accelerating fields on the current of the accelerated bunch, can be understood as the onset of a scarcely studied intermediary regime between LWFA and PWFA. This regime offers new possibilities to produce ultra-low-emittance beams as needed for next generation compact free-electron lasers (FELs)[35,36]. For instance, Manahan *et al.*[37] showed that beam loading caused by a so-called 'escort' bunch could flatten the fields experienced by a subsequently injected 'witness' beam. Alternatively, Hidding *et al.*[38] proposed to use a high-charge electron beam from LWFA as 'driver' for a subsequent PWFA stage, giving access to schemes providing electrons with ultra-low emittance[39–41] in a compact setup.

In LWFA, the wakefield is created by a high-intensity laser pulse[42–45]. Due to the *oscillating* nature of electro-magnetic waves, lasers are relatively inefficient in setting up a charge separation and a net energy gain of electrons between cycles is only possible by exploiting the gradient of the electromagnetic field. The force responsible for setting up the wakefield is thus the ponderomotive force[42]

$$\vec{F}_{\text{Pond}} = -\frac{m_e c^2}{2\langle\gamma\rangle} \vec{\nabla} \langle \vec{a}^2 \rangle.  \quad (1)$$

with $m_e$ as electron mass, $c$ is the speed of light in vacuum, $\langle\gamma\rangle$ being the cycle-averaged Lorentz factor and $\vec{a} = e\vec{A}/(m_e c) = \vec{E}_0/(m_e c^2/e) \times (\lambda_0/2\pi) \simeq 0.31 \times \vec{E}_0[\text{TV m}^{-1}] \times \lambda_0[\mu\text{m}]$ is the laser's local normalized vector potential, where $\lambda_0$ denotes the wavelength of the laser pulse and $\vec{E}_0$ stands for the electric field strength of the laser.

* a.doepp@physik.uni-muenchen.de
† stefan.karsch@physik.uni-muenchen.de

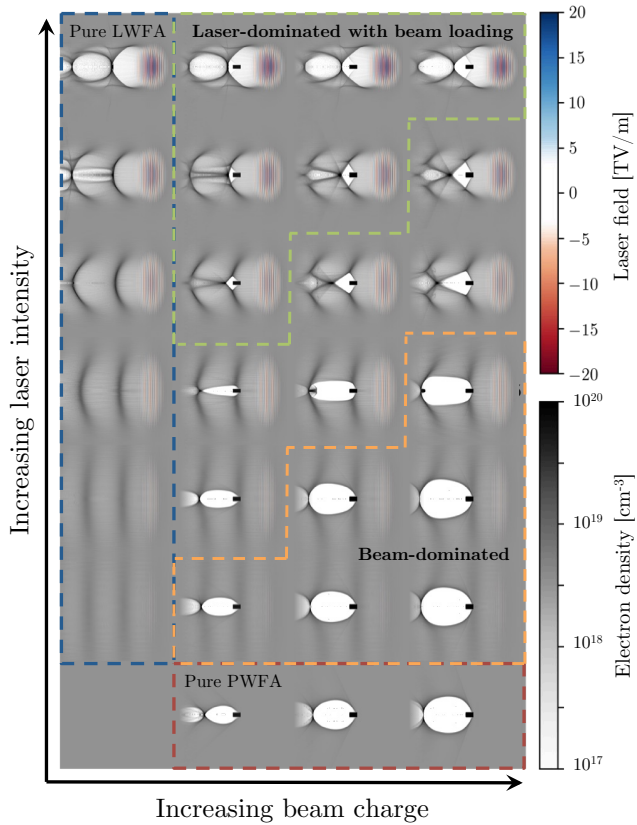

Figure 1. **Schematic view of the different regimes of laser- and beam-driven plasma wakefield acceleration** studied in the scope of this work. These plots are based on quasi-3D particle-in-cell simulations, including the transitional regimes from beam-loaded laser wakefields to beam-dominated wakefields with a weaker laser contribution. The laser intensity is varied by changing the focal spot size of a 100 TW laser pulse.

In contrast to laser pulses, particle beams used in PWFA create a *unipolar* electric space-charge, which displaces background plasma electrons much more efficiently than laser pulses. This disparity between the effect of either oscillating or unipolar fields on the plasma leads to a somewhat counter-intuitive situation. Namely, an electron beam which is accelerated in a laser wakefield can generate wakefields of similar amplitude itself, even though it carries only a fraction of the energy of the drive laser. The relation between the two driver types can be quantified in the linear wakefield regime, where the plasma perturbation introduced by a driving potential is given by[46]

$$\left(\frac{\partial^2}{\partial t^2} + \omega_p^2\right)\frac{\delta n}{n_0} = \underbrace{-\omega_p^2 \frac{n_b}{n_0}}_{\propto \phi_{\text{El}}} + \underbrace{c^2 \Delta \frac{\vec{a}^2}{2}}_{\propto \phi_{\text{Pond}}}. \quad (2)$$

Here $\omega_p$ is the plasma frequency, $n_b$ is the electron bunch density, $\delta n = n_e - n_0$, with the plasma density $n_e$ and $n_0$ denotes the background plasma density. The right hand side of the equation is composed of terms proportional to the electrostatic ($\phi_{\text{El}}$) and ponderomotive potential ($\phi_{\text{Pond}}$). Hence, both the space charge term $n_b/n_0$ and ponderomotive force term $\nabla a_0^2/2$ can have similar influence on the plasma wave formation with the limiting cases of pure LWFA ($n_b = 0$) and pure PWFA ($a_0 = 0$). Note that Eq. (2) is only valid for $n_b/n_0 \ll 1$ and $a_0 \ll 1$. For stronger drivers ($a_0 \gg 1$, $n_b/n_0 \gg 1$), the wakefield becomes non-linear and approaches the blow-out regime[47], where electrons form a thin sheath around a near-spherical ion cavity (the so-called 'bubble') behind the driver[48,49]. Most experiments however operate in an intermediary regime between perturbation and blow-out, which currently lacks a consistent theoretical description[50]. To nevertheless illustrate the interplay between both laser and particle beams which is central to this work, we use self-consistent particle-in-cell simulations[51]. As shown in Fig. 1, beam loading of electrons in LWFA affects the sheath electron trajectories, which in turn reduces the longitudinal electric fields[52]. With decreasing laser intensity this beam-loaded LWFA transitions into a regime that resembles the pure PWFA regime. Here we still observe a quasi-linear LWFA, but the plasma electron motion is dominated by the electron beam, forming an ion cavity behind the electron bunch.

While being essential to the understanding and improvement of LWFA, there exist only few experimental results on the transitional regimes indicated in green and orange in Fig. 1. First systematic studies were presented by Rechatin *et al.*[53,54] and established the correlation between charge and peak energy of an electron beam as a key signature of beam loading in LWFA. Later experiments using ionization-induced injection furthermore emphasized the impact of beam loading on the amount of charge that can be trapped inside a wakefield for a given laser power[55,56]. It has been proposed to use beam loading to flatten the accelerating fields along the injected electron bunch and thus, reduce the beam energy spread[57]. While this is mainly applicable in the dephasing-free case of electron-beam driven PWFAs, first evidence of such an optimal loading regime has been ob-served in LWFA[56]. Furthermore, indicative signs for a so-called 'self-mode transition'[58] from beam-loaded LWFA to pure PWFA have been reported[55,59–62]. In these experiments the drive laser either depletes in a long plasma target[59,60,62] or diffracts due to ionization defocusing[55], leaving the electron beam as sole driver of the wakefield.

Here we provide a comprehensive study of wakefield acceleration across different regimes of electron and laser parameters (Sec. II and III). This is followed by results on beam loading in dual-bunch configurations, which lie at the heart of escort-witness and driver-witness schemes for high-brightness beam generation (Sec. IV). In Sec. V we present first results on the transition to the beam-dominated regime using such dual-energy beams, showing the correlation of the acceleration of a witness beam with the charge of the drive beam. In Sec. VI, we provide an outlook on the scalability of LWFA towards the use of petawatt laser systems. Section VII summarizes our findings. Information on simulations and additional experimental data are given in the supplementary material.

## II. PRODUCTION OF NANOCOULOMB-CLASS ELECTRON BEAMS

The results presented in this study are based on shock-front injection, where a shock in a supersonic gas flow is used to create a sharp density downramp. The sudden change in plasma density results in a localized injection of electrons into a trapping region of the wakefield structure[63,64]. While earlier studies on shock-front injection[65–67] showed only moderate beam charges of up to few tens of pC, our recent experiments using the ATLAS-300 laser system have provided unprecedented performance regarding the stability, total charge, spectral charge density and divergence of the accelerated electron beams.

In this experiment the system provided laser pulses with 2 J energy at 27 fs duration, corresponding to a peak power of 75 TW. The laser pulses are focused by an $f/25$ off-axis parabolic mirror reaching a peak normalized vector potential in vacuum of $a_0 \approx 1.8$ at focus. A supersonic hydrogen gas jet[68] with a shock-front is placed at the laser focal spot position. The shock-front is created by the sharp edge of a silicon wafer which projects into the gas jet (cf. Fig. S2 in the supplementary material). The spectral distribution of the laser-accelerated electron bunch is characterized using a magnet spectrometer and the spectral charge density is measured by an absolutely calibrated scintillating screen[69]. Further details on all experiments are given in the supplementary material (cf. Sec. I B) including an illustration of the general setup (Fig. S1).

Figure 2 shows 100 consecutive electron spectra using the shock-front injector operated at a plateau plasma density of $n_0 = 3.0 \times 10^{18}\,\text{cm}^{-3}$. The mean charge within the peak is 338 pC and fluctuates by 11 % (36 pC rms) at a mean peak energy of 216 MeV with 4 % shot-to-shot s.d. with a rms divergence of $(0.36 \pm 0.08)\,\text{mrad}$. The peak charge is defined as the integrated charge density around the peak above a certain threshold relative to the maximum charge density. This threshold is often set to 0.5, corresponding to the full width at half maximum (FWHM). However, to study beam-plasma interaction we need a measure that reflects the whole bunch charge more accurately. We therefore use 20 % of the peak charge density as threshold unless specified otherwise. This definition contains most of the peak charge while omitting the low charge 'dark current' (see for instance Fig. S3).

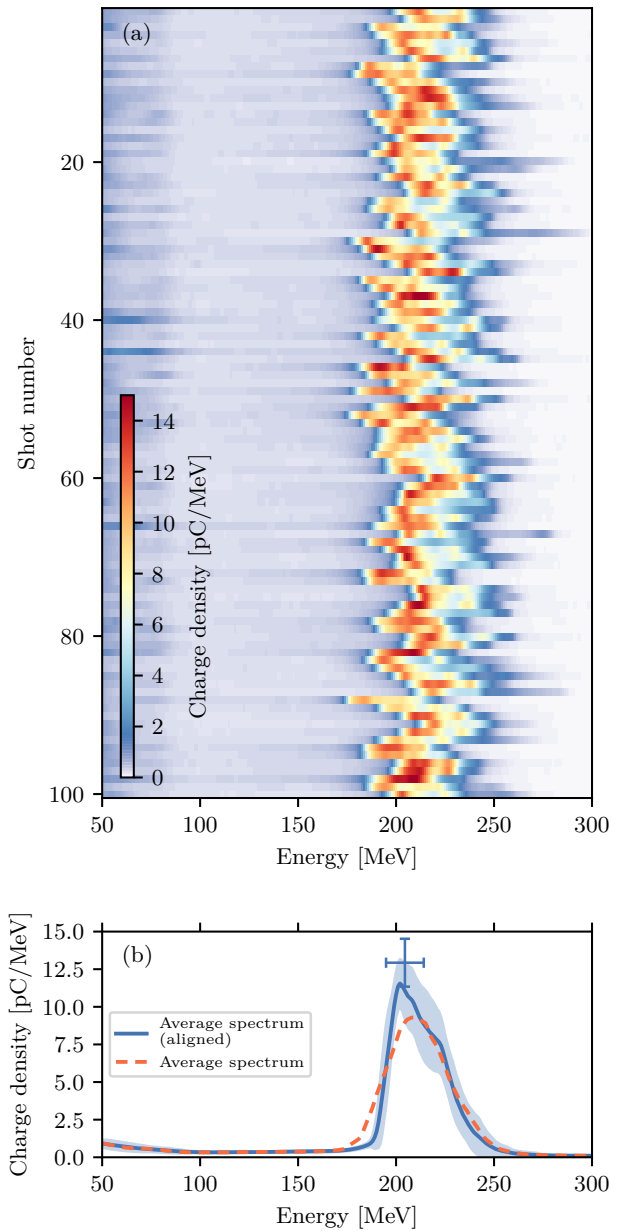

Figure 2. **Optimal performance of the shock-front injector with a 75 TW laser** (a quantitative analysis is given in the text below). (a) Angle-integrated electron spectra of 100 consecutive shots. (b) Average spectrum with and without alignment of the beams according to their central energy. The error bar marks the fluctuations in peak spectral charge density.

The FWHM absolute energy spread is $(33.0 \pm 7.2)$ MeV (rms), corresponding to a relative energy spread of 15 %. The maximal spectral charge density is $(11.7 \pm 1.4)$ pC MeV$^{-1}$, with up to 17 pC MeV$^{-1}$ for some shots. Note that the individual shots in Fig. 2(a) have a distinct spectral shape, with many of them being skewed towards lower energy. These features are lost when calculating the average spectrum since shot-to-shot fluctuations naturally lead to a normally distributed average. We have therefore realigned all the spectra according to their central energy (determined via a Gaussian fit) to preserve such features as depicted in Fig. 2(b).

Our results thus show that the amount of charge in the spectral peak in shock-front injection can be comparable to the total charge in self injection[26], with the further benefit of a clearly localized injection position. Given the known length of the gas target this permits to calculate the average accelerating fields and makes this configuration attractive for quantifying the effects of beam loading. However, the high stability of the shots shown in Fig. 2 turns into a disadvantage when studying beam loading, as they only cover a small range in charge. This is why, in the following, we deliberately concentrate on data with higher shot-to-shot charge variations, which we attribute to a less stable laser performance.

## III. EFFECTS OF BEAM LOADING ON THE ENERGY SPECTRUM

In this section we present an analysis of a data set containing 100 consecutive shots for shock-front injection at a plateau plasma density $3.5 \times 10^{18}$ cm$^{-3}$. The beam spectra, binned by the peak charge, are shown in Fig. 3(c). On average, these shots contain $(123 \pm 35)$ pC charge within the peak (28.5 % relative charge fluctuation), spanning a range from approximately 60 to 180 pC. While the average peak energy is $(232 \pm 30)$ MeV, we observe that the peak energy $E_{\text{Peak}}$ is clearly correlated to the beam charge $Q$:

$$E_{\text{Peak}}(\text{ATLAS-300}) \approx (358 \pm 35)\,\text{MeV} \\ - (1.01 \pm 0.06)\,\text{MeV} \times Q[\text{pC}]. \quad (3)$$

Additional to the reduction in beam energy at higher charges, which is a well-studied effect of beam loading[54,56], we also observe a distinctive spectral shape for beams of different charge, cf. Fig. 3(d). More specifically, there is a sharper cut-off on the low-energy side the peak, while the spectrum gradually rolls off at the high-energy side, leading to an asymmetrically distributed charge density around the peak. This effect is more pronounced for high charge bunches and similar to the average spectrum shown in Fig. 2(b). In contrast, low charge shots show a more symmetric spectrum and exhibit a quasi-Gaussian energy distribution.

To understand how the accelerated electron bunch loads the wake and affects its energy spectrum the interaction process was extensively simulated for parameters matching the ATLAS-300 experiment using the quasi-3D particle-in-cell code FBPIC[70]. To estimate the influence of different effects that could cause the observed charge fluctuations, various scans of the laser focus and power, as well as different shock heights and widths were performed (see the supplementary material for more details).

The analysis of these simulations reveals that a change in the injected charge indeed leads to a variation of the bunch spectrum due to beam loading which is similar to the experiment. In particular, the simulations show significant changes of both beam charge and energy for shifts of the focus position, as depicted exemplarily in Figure 3(e-f). The sensitivity to the focus position is due to the process of shock-front injection, which is closely related to the intensity of the laser at the downramp. This is because the area in the phase space where charge can be trapped depends on the radius $r$ of the ion cavity ('bubble radius'), which scales as $r \propto \sqrt{a_0/n_e}$, see Ref.[71]. Hence, the normalized laser intensity directly influences the injection process and the involved accelerating fields. Notably, as shown in Fig. S9, the injection is even sensitive to focus position shifts on a scale of 0.2 mm. This is well below the Rayleigh length of 1.1 mm and the equivalent amount of global wavefront curvature in the near field is $\Delta\lambda \lesssim \lambda/10$. These kind of shifts can easily result from shot-to-shot wavefront fluctuations and the results thus highlight the need of wavefront stabilization to increase the stability of this type of injector.

The simulations also highlight two effects which can cause the observed asymmetry of the electron spectrum, namely the current profile and beam loading. The former is a direct consequence of the injection process which favors electrons to be injected at the very beginning of the downramp (cf. Fig. S8). These electrons constitute the front of the electron bunch leading to an asymmetric current shape. However, the simulations in Fig. 3(e-f) show that the current profile alone does not account for the degree of asymmetry observed in the energy spectrum. Instead, we attribute it to beam loading.

Due to the tight spatial and temporal confinement of the injection, the electrons' final energies only depend on the strength of the local accelerating field. Therefore electrons near the back of the bubble gain more energy in case of low charge, and thus weak beam loading, due to the prevailing stronger fields there. Consequently the phase space of such electron bunches exhibit a positive quasi-linear chirp. However, the influence of the bunch's space charge on the wakefield changes the acceleration gradient and results in the bunch acquiring a nonlinear chirp in phase space (curvature of phase space in Fig. 3(e)). Being essentially the projection of this phase space onto the energy axis, the electron spectrum therefore exhibits an asymmetry, i.e. the signature of the nonlinear phase space chirp is a skewed energy spectrum (more details and a quantitative analysis of the skewness for each individual spectrum in Fig. 3 can be found in the supplementary material).

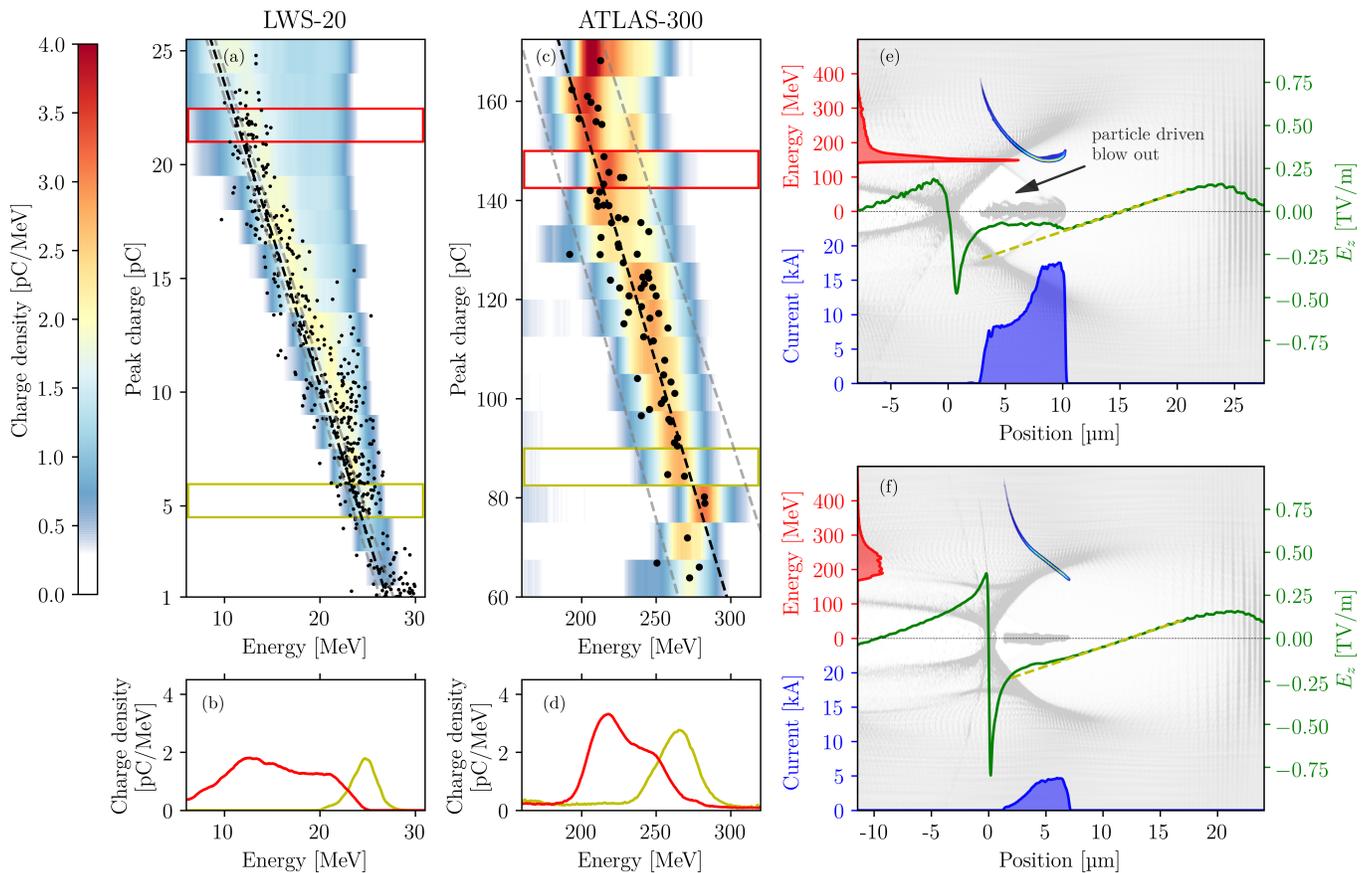

Figure 3. **Beam loading effects on single laser-accelerated electron bunches.** Change of the spectral shape with beam charge. (a) binned LWS-20 data including the location of the peak energy and peak charge for the individual shots. Line-outs for two bins of high and low charge as marked in their respective colors in (a) are given in (b). (c) shows binned spectra of a data set acquired using ATLAS-300. Two individual line-outs are given in (d). Dashed black and gray lines indicate linear fits and standard deviations according to Eq. (4) and (3). Figures (e) and (f) summarize the results of particle-in-cell simulations for the ATLAS-300 parameters for different focus positions, yielding either high beam charge (e) or low beam charge (f). The on-axis field is plotted in green, with dashed yellow lines as linear fit to the front of the wakefield to show the influence of beam loading. The beam current is plotted as shaded blue area (bottom), while the spectrum is shown on the left (shaded red). The grey background shows a map of the electron density, while the colored plot shows the electron phase space.

As beam loading effects increase with the total charge contained in the accelerating cavity we expect the asymmetry to become more pronounced for higher total beam charges which clearly matches the measured data. Indeed, for the ATLAS-300 data the calculated skewness for almost each electron spectrum is positive and increases with charge, in accordance with simulations. The positive skewness is a typical feature of beam loading in shock-front injection. Our simulations suggest that its dependence on the injected charge can be used as an indicator for the onset of beam loading. It is important to notice that this signature gets easily lost when a beam with large divergence and a non-imaging spectrometer is used. Future studies including current measurements with coherent transition radiation (CTR) diagnostics[61] may even allow for quantitative measurements of beam loading or full reconstruction of the electron beam phase space, provided the time ambiguity of the CTR measurement can be resolved.

To show that the effects discussed in this section are general signatures of beam loaded LWFA and not restricted to our particular regime of operation, we have also analyzed a data set on shock-front injection from the LWS-20 OPCPA system in its former incarnation at the Max-Planck-Institute for Quantum Optics (MPQ). This few-cycle laser delivered 8 fs, 65 mJ pulses as presented in Schmid *et al.*[65]. Figures 3(a-b) are based on data obtained at a plateau plasma density of $\sim 3.1 \times 10^{19}\,\mathrm{cm}^{-3}$. The data set consists of 490 electron beam spectra with more than 1 pC peak charge each, with an average peak charge of $(10.5 \pm 6.0)$ pC and an average peak energy of

$(21.0 \pm 6.4)$ MeV. As for ATLAS, we measure a correlation between peak energy and charge, in this case

$$E_{\text{Peak}}(\text{LWS-20}) \approx (29.0 \pm 0.1)\,\text{MeV} \\ - (0.84 \pm 0.01)\,\text{MeV} \times Q[\text{pC}]. \quad (4)$$

Furthermore, the data show a very similar behavior regarding the spectral shape at low and high beam charges. The striking similarity between these two very different experiments underlines a fundamental feature of beam loading, namely that it is not a purely charge-driven effect occurring in nanocoulomb-class LWFA. Instead, it is the result of an interplay between the drive laser, its wakefield and the accelerated electron beam. Indeed, Tzoufras et al.[72] showed that beam loading effects in the blow-out regime depend on the laser intensity and the wakefield amplitude. Here, a constant accelerating field $E_s$ is generated for a charge $Q_s$ and a peak potential $a_0$ with $Q_s \propto a_0^2/E_s$. Assuming that we are close to a loaded regime and far from dephasing, we can estimate $E_s$ as the ratio between acceleration distance and beam energy, yielding $E_s(\text{LWS-20}) = 0.2\,\text{TV\,m}^{-1}$ and $E_s(\text{ATLAS-300}) = 0.066\,\text{TV\,m}^{-1}$. Regarding $a_0$, simulations of ATLAS-300 show that the value is close to the matched parameters for the plateau density and laser power, i.e. $a_0(\text{ATLAS-300}) \approx 3.7$. In contrast, in the LWS-20 case the total plasma length of $< 0.2\,\text{mm}$ is insufficient to reach an equilibrium between diffraction and self-focusing. From simulations we can estimate that $a_0$ varies from $a_0(\text{LWS-20}) \approx 1.5$ at the shock up to 3.0 at the exit. The ratio between the beam loading charge in both cases is therefore expected to be of the order $Q_s(\text{ATLAS-300})/Q_s(\text{LWS-20}) \approx 5-20$, which is compatible with the experimental observations. Hence, even though the LWS-20 experiment produced an order of magnitude less charge, the two experiments reach a similar regime of beam loading. This means that the results presented in this work are not only applicable to our specific experimental conditions, but that similar behavior is expected to occur in any LWFA operating in a comparable regime of beam loading.

Last, it should be noted that the measured charge-energy correlations from Eq. (4) and Eq. (3) show that the electron peak energy fluctuations are dominated by a charge-dependent component. This observation emphasizes the importance of reducing charge fluctuations in LWFA, as shot-to-shot variations of the beam charge will directly translate into peak energy fluctuations. The latter are particularly detrimental to applications relying on chromatic beam transport elements, such as free electron lasers[73]. Given the sensitivity of shock-front injection on the laser intensity at the density transition, this will likely require the development of active wavefront stabilization.

## IV. BEAM LOADING WITH DUAL-ENERGY ELECTRON BEAMS

So far, we have considered how a high-current electron beam modifies its own accelerating field and thus, its final energy spectrum. However, such a beam will not only affect its direct vicinity, but also modify the wakefield formation in its trail. Here, we are going to discuss how beam loading of a first injected electron beam will affect secondary electron beams. As discussed in the introduction, this kind of scenario is particularly promising due to various proposed advanced injection schemes relying on similar conditions. In the following we will discuss two different schemes used for injection of multiple electron beams, namely combined shock-front and colliding-pulse injection into the same wakefield cavity[74], as well as shock-front injection into subsequent wakefield periods[75]. Due to the different location of the bunches in both schemes (intra-cavity versus inter-cavity), these methods are complementary and can be used as a basis for studying escort-witness and driver-witness configurations.

In the first case, we use a gas jet with shock-front injection and split off a part of the main laser beam to obtain a second pulse propagating in the opposite direction and containing an energy of 0.3 J. This counterpropagating pulse is then focused onto the gas target reaching a normalized peak intensity of up to $a_1 \simeq 0.9$. When this colliding pulse is active, we can inject a second bunch with lower energy and charge into the same wakefield period via optical injection. Both bunches' peak energies are plotted in Fig. 4. As in the previous section, the energy of the first bunch depends on its own charge $Q_1$ ($E_1 \approx (387 \pm 4)\text{MeV} - (1.3 \pm 0.1)\text{MeV} \times Q_1[\text{pC}]$). In addition, we now observe a similar trend for the optically injected beam whose energy also correlates with the charge $Q_1$ of the *first* bunch ($E_2 \approx (175 \pm 5)\text{MeV} - (0.7 \pm 0.1)\text{MeV} \times Q_1[\text{pC}]$), cf. Fig. 4(a). In contrast, no statistically significant correlation between the charge of the second bunch $Q_2$ and the energy of either bunch $E_1$ is found (Fig. 4(b)). The first bunch therefore clearly influences the second bunch, which in this case sits in the same plasma period. Nevertheless, the data shows that beam loading effects of the trailing low energy bunch on itself are negligible.

In the second experiment, we just used a single laser pulse with shock-front injection but optimized the laser and target parameters (focus and shock position, etc.) to yield a second beam at lower energies. Such trailing beams have already been observed previously (see for instance Ref.[74,76]), but usually with much lower charge than the main beam. These trailing beams stem from shock injection into the second plasma period and are therefore intrinsically separated from the leading beam. This bunch distance is approximately given by the plasma wavelength $\lambda_p$. Again, the energy of the leading beam correlates with its own charge, even though weaker than in the previous case ($E_1 \approx (259 \pm 3)\text{MeV} -$ $(0.11 \pm 0.03)$MeV $\times Q_1$[pC]), because the target parameters are slightly different from the previous experiment. However, in contrast to previous studies[54], the second beam shows no correlation beyond the statistical error ($E_2 \approx (113 \pm 3)$MeV $- (0.02 \pm 0.02)$MeV $\times Q_1$[pC]), (Fig. 4(d)). Furthermore, we now observe that the energy of the second beam correlates with its own charge ($E_2 \approx (113 \pm 3)$MeV $- (0.18 \pm 0.06)$MeV $\times Q_2$[pC]). Even more, when the charge of the second beam increases, the energy of the first beam tends to be higher ($E_1 \approx (231 \pm 4)$MeV $- (0.28 \pm 0.07)$MeV $\times Q_2$[pC]), as well, (Fig. 4(e)).

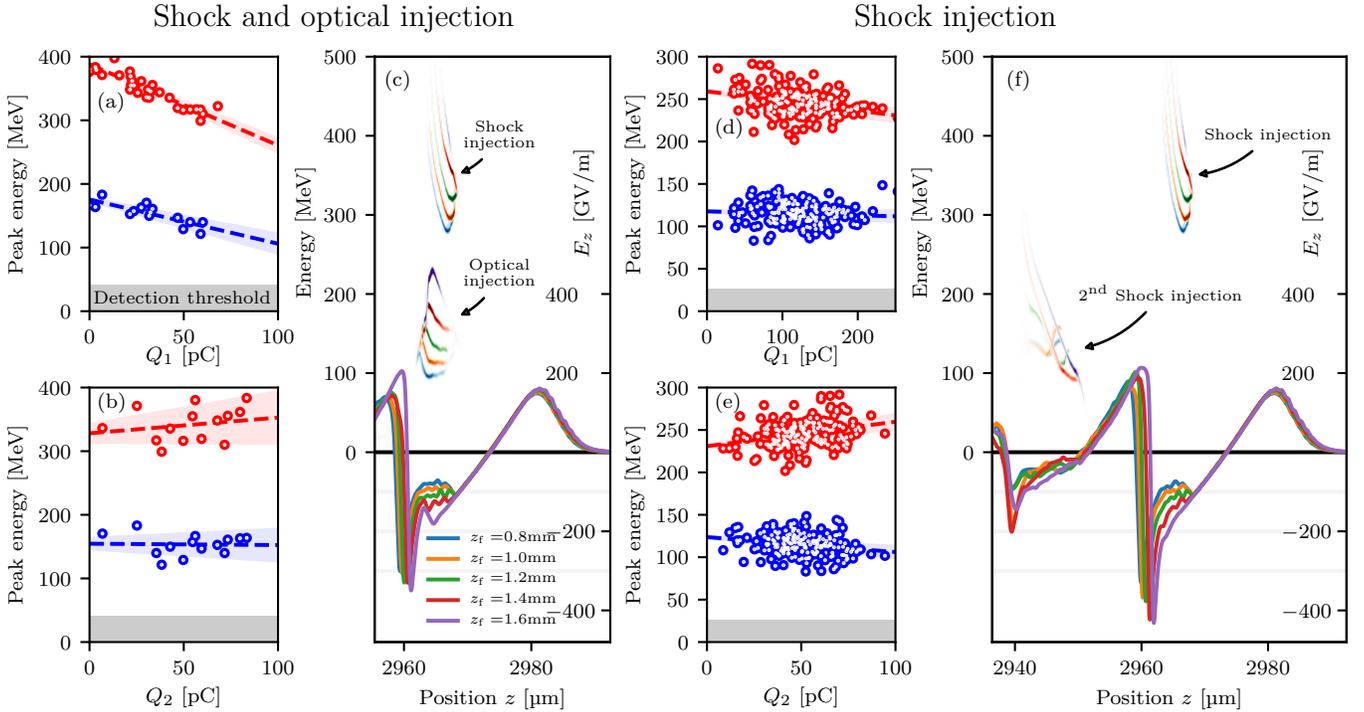

Figure 4. **Effect of beam loading on dual-energy electron beams** injected into either the first or second wakefield cavity. *First Column:* Experimental data for electron injection into the first wakefield period via shock and optical injection. The data is either sorted by the charge in the high energy beam (denoted with charge $Q_1$) (a) or by the charge in the low energy beam ($Q_2$) (b). Dotted lines show linear fits and the shaded areas indicate the corresponding confidence interval. The data show a clear correlation between the charge of the shock-injected beam $Q_1$ with the energy of the optically injected beam $E_2$ (a). (c) Results from particle-in-cell simulations showing both the phase space and the longitudinal on-axis fields. Beams of different charge are injected by changing the absolute focus position $z_f$ relative to the density transition $z_{shock} = 0.8$ mm. Higher charges $Q_1$ decrease the accelerating fields in the wakefield via beam loading leading to a lower energy of the trailing optically injected electron bunch. *Third Column:* Experimental data for electron injection into the first and second wakefield period via shock injection. The high energy beam seems to have no influence on the particle energies in the second wakefield period (d). In contrast to (b), the charge of the second bunch ($Q_2$) is correlated to the energy of the first electron bunch (e). (f) Particle-in-cell simulations with same color coding as in the aforementioned plot showing the same behavior as in (d). The energy of the second beam is uncorrelated to the charge in the first beam which matches the experimental data.

To understand this behavior, we have simulated the two scenarios using particle-in-cell simulations. For better comparison, the parameters of these simulations are modelled on the experiment with shock-front injection into two buckets. This way injection into the second bucket can be switched on and off by adjusting the simulation window length while keeping the evolution of the plasma wave identical. Based on our previous findings, we use the focus position to adjust the injected charge while keeping the wakefields similar. This results in electron spectra with decreasing charge and increasing energy the further away the focus position lies from the shock at $z_{shock} = 0.8$ mm. For colliding-pulse injection we initialize a second counterpropagating laser beam with $a_1 = 0.5$ at a distance $\Delta z = 1.1$ mm behind the shock on the density plateau. The phase space of the electrons injected due to this optical injection process shows a clear dependence with the scanned focus positions $z_f$ and hence, charge of the first beam (cf. Fig. 4(c)). Looking at the longitudinal on-axis wakefield $E_z$ we can identify the cause of this correlation to be the different amount of beam loading induced by the first beam, i.e. the differing amplitude of $E_z$ for each case. This result confirms the hypothesis deduced from the experimental findings, namely that the second bunch is a direct witness of beam loading induced by the first. The simulations also indicate that the charge injected via colliding-pulse injection is reduced due to beam loading of the first bunch.

For the second case, we use a longer simulation window to observe also injection into the second wakefield period (cf. Fig. 4(f)). As in the experiment, the energy of the second injected bunch does not seem to be determined by the charge and beam loading induced by the first bunch, even though it clearly perturbs the first wakefield. In fact, for all different cases the slope of the wakefield in the second period appears to be nearly independent of beam loading from the first bunch and only changes due to beam loading of the second bunch itself (explaining the correlation $E_2|Q_2$, cf. Fig. 4(f)).

The simulations also help us to understand the $E_1|Q_2$ correlation, which is actually caused by a relation between $Q_1|Q_2$ and $Q_1|E_1$: During any downramp injection process, electrons are more likely to be injected into subsequent wakefield periods due to the progressively slower phase velocity. This means that as less electrons are captured by the first wakefield period, more electrons can be trapped in the second period and vice versa. We have already established how the charge of the first beam affects its energy, hence explaining the observed correlation.

In accordance with quasi-3D PIC simulations, our results therefore demonstrate how charge and energy of multiple electron bunches are correlated. In particular, our experimental measurements on bunches in subsequent wakefield periods show an unexpectedly weak correlation between the charge of the first bunch and the energy of the second, an effect which could be reproduced in PIC simulations for our conditions. In future experiments this method could be extended to probe local fields and effects of beam loading across subsequent wakefield cavities, and hence yield quantitative properties of laser wakefields under different conditions.

Moreover, optical injection can provide an electron beam that witnesses the effect of self-fields of another beam inside the *same* cavity. This configuration can be used for producing a witness-escort pair as proposed by Manahan *et al.*[37] in the context of PWFA. Here the shock-injected beam could potentially act as an escort and flatten the LWFA field that is experienced by an independently injected witness bunch in order to reduce its energy spread. Indeed, this effect is clearly visible in the simulations. For instance, if the laser is focused onto the shock at $z_f = 0.8$ mm, shown in Fig. 4(c), a higher charge $Q_1$ is injected, which leads to smaller energy spread in phase space of the optically injected bunch. So far, the experimental data are not conclusive about this, but it is planned to study this configuration further in future campaigns.

## V. TRANSITION TO THE BEAM DOMINATED REGIME

In the preceding sections we have established that shock-injected electron beams have a sufficiently high charge density to modify the electric fields of the laser wakefield and thus influence both their own spectrum and the spectrum of other beams. The underlying force is fundamentally the same that is responsible for the excitation of plasma wakefields, and as visible in Figure 1, even in the laser-dominated regime the electron beam generates its own wake, i.e. expelling any electrons that were not blown out from the bubble by the ponderomotive force of the laser. As the laser gets weaker in intensity, e.g. due to energy depletion or diffraction, the electron beam will become the dominant driver of the wakefield, see lower part of Figure 1. Indeed, it was recently observed that a laser-generated electron beam can drive plasma waves in a pure PWFA regime[77].

While this effect can occur naturally in form of a self-mode transition, it is not easy to control and offers little insight into the transition process. The most systematic studies in this regard used gas targets of variable length[61,62], but their interpretation relies heavily on simulations because the laser intensity cannot be controlled independently of the electron beam energy in this case. In particular, due to self-guiding of the laser pulse, the self-mode transition will occur close to laser depletion at a distance $L_\mathrm{depl} \sim (n_\mathrm{cr}/n_\mathrm{e})c\tau$, with $c\tau$ being the pulse length[71]. For resonant wakefield excitation this value is close to the dephasing length $L_\mathrm{deph} \sim (n_\mathrm{cr}/n_\mathrm{e})(2/3)R_\mathrm{bubble}$, where $R_\mathrm{bubble}$ is the bubble radius[71]. This means that the first electron beam will also lose energy due to dephasing, which is both inefficient and makes it difficult to distinguish the LWFA- and PWFA-dominated cases in this situation.

To circumvent these problems we follow another approach here. Instead of extending the gas target towards self-mode transition we use the different divergence of both the electron and laser beam, i.e. $\theta_\mathrm{elec} \sim 1$ mrad for the former versus $\theta_\mathrm{laser} \gtrsim 40$ mrad for the latter[78], to strongly reduce the laser intensity after a vacuum gap. We can then place a second gas target behind the first one such that the electron beam takes over as the primary wakefield driver. Importantly, in this setup the electron beam energy remains unchanged during the transition from the beam-loaded LWFA to the beam-dominated regime in the second stage.

A distinctive sign for the transition to the beam dominated regime is that a second witness bunch with the correct delay will experience a positive correlation with the first bunch's charge due to the increasing wakefield amplitude induced by a higher driver bunch charge. As discussed before, dual-bunch shock injection intrinsically provides a driver witness pair with a bunch separation of $\sim \lambda_\mathrm{p}$. The leading $200-300$ MeV bunch serves as driver of the wakefield in the second nozzle, whereas the subsequent $75-125$ MeV bunch probes these fields. Please recall that the energy of the low energy beam is essentially independent of the charge of the high energy beam in this case. We interferometrically measure a peak plasma density of $1.4 \times 10^{18}$ cm$^{-3}$ (i.e. $\rho \simeq 0.225$ pC µm$^{-3}$) in the second plasma from a nozzle with an opening diameter of 1 mm. For this density we expect the second bunch to be situated in the accelerating phase of the beam-dominated

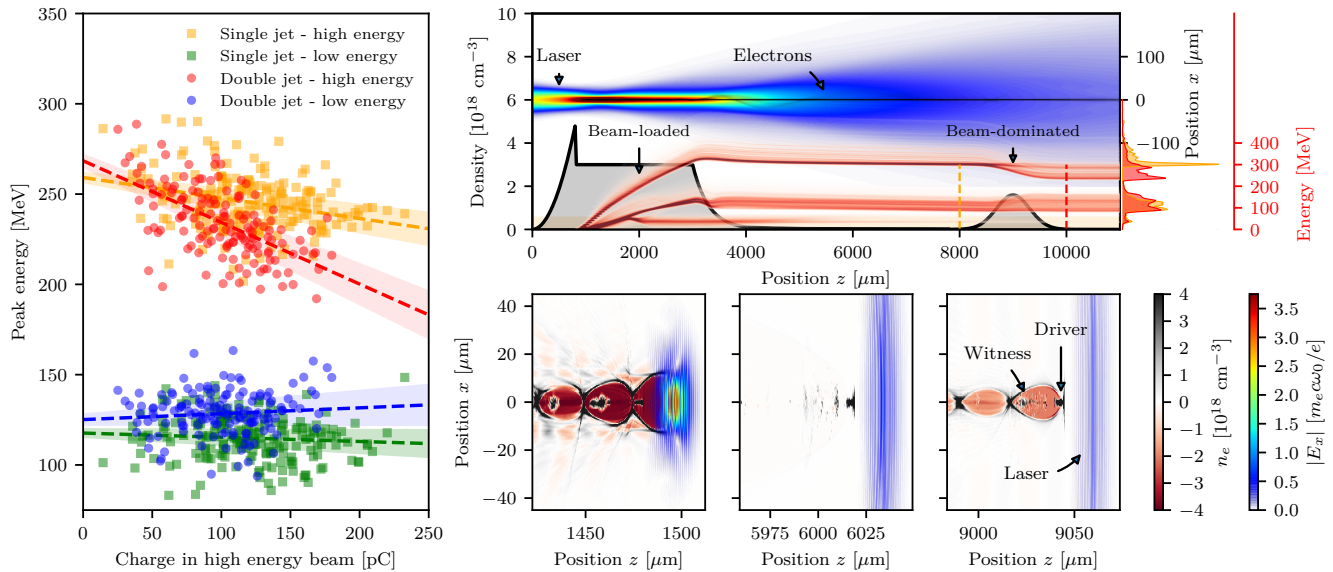

Figure 5. **Electron acceleration in the beam-dominated regime.** *Left:* Energy from the driving electron bunch (high energy) is transferred to the witness bunch (low energy) in the second stage. The higher the drive bunch charge, the higher the witness energy gain. The driver itself therefore loses energy and charge. *Top Right:* Results of a start-to-end simulation showing both the laser and beam evolution (upper part), as well as the plasma density (grey, lower part) and energy spectrum at each position (red). The leading electron beam contains a charge of 105 pC. It gets decelerated in the second jet, while a part of the second bunch is accelerated, just as observed in the experiment. The energy spectrum before the jet at $z = 8$ mm and after the jet at $z = 10$ mm is shown on the right. *Bottom Right:* Snapshots for three different positions, i.e. inside the laser wakefield accelerator, during propagation in the near-vacuum between both jets and inside the beam-dominated second jet. The respective plasma densities and electric fields are color coded.

wakefield.

As presented in Figure 5, the beam parameters are significantly altered by the second stage. Similar to Chou *et al.*[79], we also observe that the average beam charge is reduced by about 20 % and the high energy beam is decelerated. While this alone could be explained with dephasing in a laser-driven wakefield, we also observe a *charge-dependent acceleration* of the witness by 10 MeV to 20 MeV, translating into an acceleration gradient of $10\,\mathrm{GeV\,m^{-1}}$ to $20\,\mathrm{GeV\,m^{-1}}$ (see Fig. 5). As mentioned earlier, this is the opposite of what would be expected in the laser-dominated case.

We have simulated the entire target setup in start-to-end simulations using a boosted frame[80,81]. The results are depicted in Fig. 5. According to simulations the normalized vector potential of the laser drops from $a_0 \simeq 3.2$ at the exit of the first gas jet to below $a_0 < 0.6$ when entering the second jet due to diffraction. There is no beam collapse due to self focusing in the second jet and hence the laser drives only a weak linear plasma wave.

In contrast, the small divergence of $\sim 1$ mrad keeps the density of the electron bunch high and allows them to become the primary driving force of the wakefield. The simulations clearly show deceleration of the high energy bunch in the second jet, combined with acceleration of the low energy beam. As in the experiment, this effect depends on the charge of the first beam. In simulations we mimic this effect by adjusting the focus position (cf. Sec. IV), where the case displayed in Fig. 5 corresponds to a focus at $z_f = 1.0$ mm with a total charge of 105 pC in the driver. Here the observed acceleration and deceleration is comparable to the experiment. An example for the case with higher charge in the driver (125 pC) is given in the supplementary material. In this scenario the spectra of high and low energy beams almost overlap after leaving the second gas target.

While the electron beam clearly dominates the wakefield formation, the head of the bunch still experiences the linear wakefield driven by the laser. With appropriate timing a weak laser pulse could therefore, for instance, be used to mitigate head erosion in laboratory-scale PWFA experiments. Expanding on this idea, the laser can also be used for guiding of the electron beam in a configuration similar to a laser-plasma lens[82]. In the limit of negligible laser intensity this setup becomes equivalent to a compact externally seeded PWFA stage such as presented by Kurz *et al.*[75]. However, even for low laser intensity the configuration discussed here has the advantage that no plasma mirror is necessary to block the laser and that the laser pre-ionizes the plasma.

## VI. TOWARDS JOULE-CLASS ELECTRON BEAMS

The results presented so far have been obtained with sub-100-TW lasers. With higher laser power, the accelerated charge is also increasing and the energy contained in the relativistic electron beam is approaching the joule energy level.

To quantify the scaling behavior, we have performed shock-front injection with various laser energies and a pulse duration of $\sim 27\,\text{fs}$. In a first scaling test from 10 TW to 70 TW we found that the charge in the peak follows an approximately linear trend

$$Q_\text{Peak} \approx 5.5\,\text{pC} \times (P[\text{TW}] - P_\text{inj}[\text{TW}]), \quad (5)$$

where $P_\text{inj} = 15\,\text{TW}$ is the power threshold for injection. The laser power in these experiments was adjusted by detuning the pump lasers and was generally limited by the damage threshold of the compressor gratings to below 70 TW. Nevertheless, in a separate campaign, few shots were taken with 110 TW (3 J) on target. Here we observed even more charge with up to 1.2 nC in the energy range above 80 MeV and a peak charge of 602 pC, corresponding to 50 % of the total charge, at a maximal peak charge density of $15\,\text{pC\,MeV}^{-1}$ (see Fig. S3). These data are still in agreement with the above linear scaling. The total energy of the electron beam observed on the spectrometer is 288 mJ, corresponding to an energy transfer rate from the laser pulse to the electron beam of almost 10 % (for further details see Sec. I B in Supplementary Material).

As part of the commissioning of our new ATLAS-3000 multi-petawatt laser we have also performed first tests with an energy on target of up to 10 J and a peak power reaching 330 TW[83]. A typical shot from these experiments is shown in Fig. 6, also showing the skewed spectral shape established in Sec. III. Over a series of 50 shots we measure a charge in the peak of $(1.4 \pm 0.2)\,\text{nC}$, $(1.1 \pm 0.2)\,\text{nC}$ in FWHM, with an average peak charge density of $(22 \pm 4)\,\text{pC\,MeV}^{-1}$ and an average energy of 334 MeV. Integrating over the entire spectrum, we measure a total transferred energy to the relativistic electrons of $(0.6 \pm 0.1)\,\text{J}$. Notably, the measured beam charge remains below the prediction of the scaling (1.7 nC, cf. Eq. (5)).

In order to extrapolate towards the petawatt-level we have therefore performed PIC simulations for different laser powers. The simulations are in agreement with the experimental data and show a transition from a linear scaling $Q_\text{Peak} \propto P$ to $Q_\text{Peak} \propto \sqrt{P}$ for $P \gtrsim 200\,\text{TW}$, see supplementary material Fig. S4. We interpret this behavior as a transition from the partially-loaded to the fully-loaded laser wakefield regime[55], with the latter having a well-established $\sqrt{P}$ scaling[71]. Based on our analysis, we estimate a peak charge of the order of 2.5 nC and more than 100 kA peak current for a one petawatt laser driver, which is more than twice the value predicted by the scaling for self-truncated ionization injection[56].

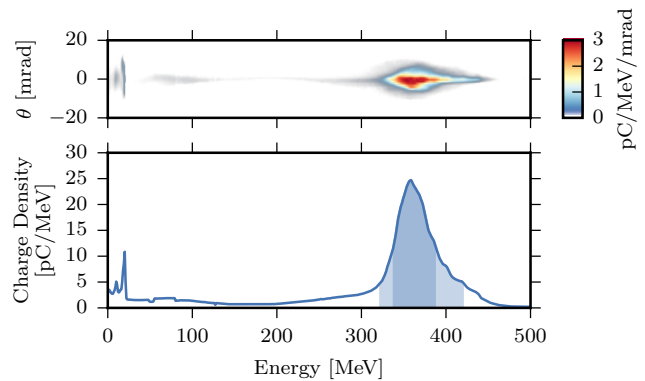

Figure 6. **Typical shot from shock-front injection using a 300-TW laser.** The peak contains a charge of 1.3 nC (semi transparent shaded area) (0.95 nC in FWHM, shaded area).

## VII. CONCLUSIONS AND OUTLOOK

We have presented results on nanocoulomb-class electron beams produced via LWFA, which feature exceptionally high spectral charge densities and low divergence. While the beam parameters are promising for demonstrating a laser-based free electron laser in the near-optical regime[84], many other applications require even better beam quality in terms of emittance and, more generally, brightness. Due to their high charge density, the laser-accelerated beams can be used to access new regimes of beam-loaded and beam-dominated wakefield acceleration. This intermediate regime of LWFA and PWFA holds promise for the generation of ultra-low-emittance beams in a compact setup in the future.

Our analysis clearly shows that we are capable of reaching different regimes of beam loading, evidenced by a dependence of both the final beam energy as well as the spectral shape of laser-accelerated electron beams on the beam charge. This effect was consistently observed in different regimes of LWFA, i.e. in the few-cycle regime using the 100-mJ-class LWS-20 laser and in the joule-class regime with the ATLAS-300 laser.

When combined with not a single, but two electron bunches, beam loading can become a powerful tool to optimize LWFA. In a proof-of-principle experiment we have obtained first results on beam loading for electron beams injected into the same wakefield period as well as subsequent wakefields. The former serves as precursor for a escort-witness configuration[37], while the latter was used to study beam-dominated LWFA in a driver-witness configuration[38]. Here we demonstrated a charge dependence of the witness acceleration in a second wakefield stage, a clear signature for the transition to the beam-dominated regime. The energy gain in this setup was mainly limited by the energy separation between both bunches, which will be increased in futures experiments. Furthermore, bunch duration and separation measurements via coherent transition radiation are expected to yield more quantitative information on both dual-beam configurations.

As an outlook, the scalability of the shock-injection scheme was studied by means of experiments and simulations in the ranges up to 330 TW and 1 PW, respectively. Based on these studies it is likely that laser wakefield accelerators driven by petawatt lasers will very soon deliver joule-class electron beams with 100-kA-class currents, which will give access to new regimes of laser-beam interaction and radiation generation.

## ACKNOWLEDGMENTS

We acknowledge K. Schmid and C. M. S. Sears who performed experiments on the LWS-20 laser at MPQ.

This work was supported by the DFG through the Cluster of Excellence Munich-Centre for Advanced Photonics (MAP EXC 158), TR-18 funding schemes, by EURATOM-IPP (ENR-IFE19.CCFE-01) and the Max Planck Society. The authors gratefully acknowledge the Gauss Centre for Supercomputing e.V. (www.gausscentre.eu) for funding this project by providing computing time on the GCS Supercomputer SuperMUC at the Leibniz Supercomputing Centre (www.lrz.de). L.V. acknowledges the support from Vetenskapsrådet / Swedish Research Council (2016-05409 and 2019-02376).

## Author contributions

J.G. and A.D. contributed equally to this work. A.D., J.G., M.G., H.D., S.S., G.S., and S.K. set up and/or performed the experiments with ATLAS-300. M.F., A.D., J.G., M.G., H.D. and S.K. performed the experiments with ATLAS-3000. A.D. and J.G. analyzed the data from ATLAS-300 and performed simulations. A.B. and L.V. provided complementary data with LWS-20 results. M.F. analyzed data from the ATLAS-3000 commissioning experiments. All authors discussed the results. A.D. and J.G. wrote the paper. S.K. supervised the project.

# Supplementary material: Electron acceleration in beam-loaded and beam-dominated laser-plasma wakefields

J. Götzfried,[1,2] A. Döpp,[1,2] M. Gilljohann,[1,2] M. Foerster,[1] H. Ding,[1,2]
S. Schindler,[1,2] G. Schilling,[1] A. Buck,[2] L. Veisz,[2,3] and S. Karsch[1,2]

[1]*Ludwig-Maximilians-Universität München,*
*Am Coulombwall 1, 85748 Garching, Germany*
[2]*Max-Planck-Institut für Quantenoptik, 85748 Garching, Germany*
[3]*Department of Physics, Umeå University, SE-901 87, Umeå, Sweden*

This supplemental material contains additional information on the experimental setup, discusses experiments on the power scaling of shock-front injection and elaborates on the definition of skewness for electron spectra obtained in the experiments. Furthermore, we provide detailed information on the configurations used for the various simulations performed throughout the manuscript, as well as additional simulations on the origin of shock-injected electrons, the behavior of shock-front injection with respect to the laser focal plane, the self-mode-transition regime and start-to-end simulations with different amounts of injected charge.

**CONTENTS**

# I. SUPPLEMENTAL EXPERIMENTAL DATA

## A. Experimental setup

If not stated otherwise, experiments were performed at the Laboratory for Extreme Photonics (LEX) in Garching, Germany, using the ATLAS Ti:Sa laser system. The laser delivers 800 nm laser pulses with 27 fs duration and up to 3 J to the target chamber as shown in Fig. S1. There, an off-axis parabolic mirror is used to focus the laser to an estimated peak intensity of $6.9 \times 10^{18}\,\mathrm{W\,cm^{-2}}$ corresponding to a normalized vector potential of $a_0 \approx 1.8$.

Data for shock-front injection with 330-TW laser pulses were taken with the upgraded ATLAS laser, now at the Center for Advanced Laser Applications (CALA) in Garching. Here the laser delivered 800 nm laser pulses with 27 fs duration and up to 10 J to the target chamber, where an off-axis parabolic mirror is used to focus the laser in an approximately $f/20$ configuration to a spot size of 20 µm (FWHM).

As gas target we use a supersonic de Laval nozzle with 5 mm orifice for the first jet and a 1-mm-de Laval nozzle for the second stage (Sec. IV and V). The density profile in both cases is retrieved via interferometry and cross-checked with few-cycle shadowgraphy[1], see Fig. S2.

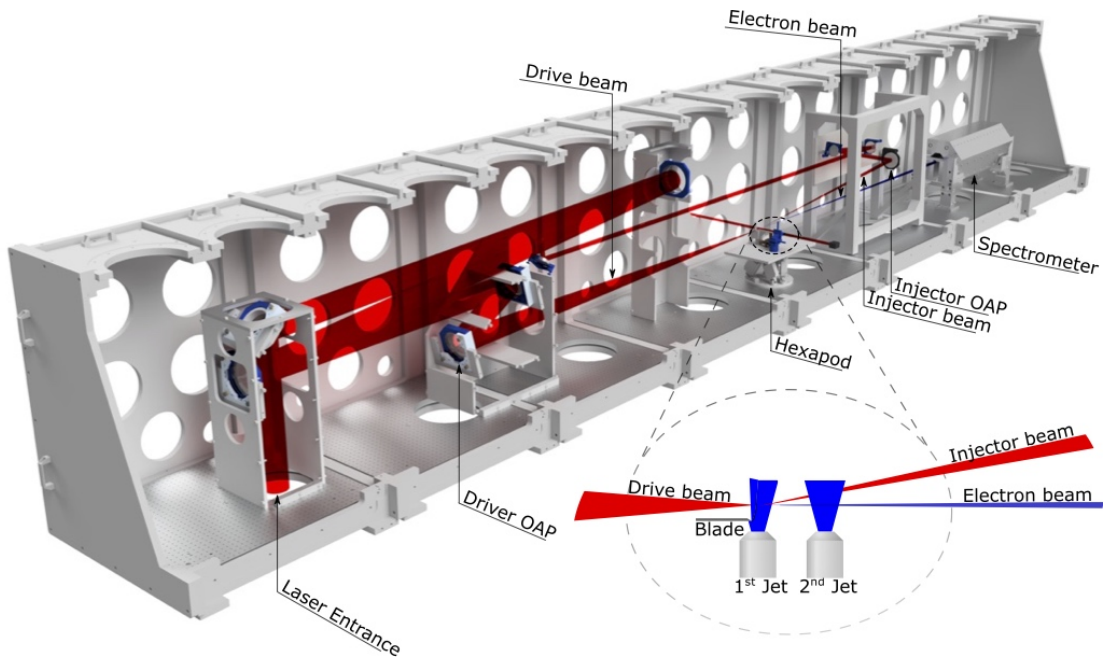

Figure S1. Experimental setup. The laser is first delayed in order to allow synchronization with the other beams (optical probe beam and the injector beam) and then focused by an off-axis parabolic mirror (OAP) onto the target. The target consists of one (Sec. II/III/VI) or two gas jets (Sec. IV/V), where a blade is used to perturb the gas flow and induce a downward density-step in a shock front. Optionally a part of the main pulse is clipped out of the drive beam and used as injector beam for colliding pulse injection.

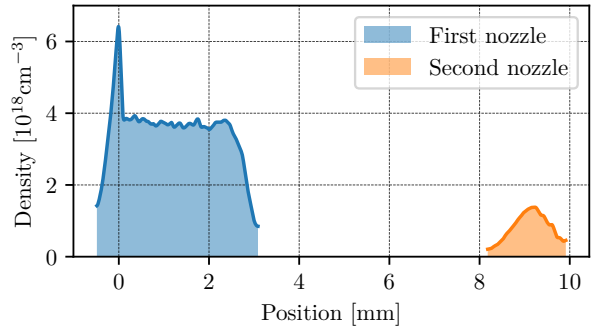

Figure S2. Measured density profile of the double nozzle setup used for PWFA studies. The first acceleration stage consists of a de Laval nozzle with a 5 mm orifice. The gas flow here is perturbed to form a sharp density downramp (shock). The second stage roughly located 7.5 mm behind the first nozzle (center-to-center) has an orifice of 1 mm and is operated at lower hydrogen pressures.

In both cases a 0.8 m long 0.85 T dipole magnet is used to disperse the electron beam, which is then hitting a scintillating screen. The scintillation signal from the electrons is compared to the signal of a calibrated tritium capsule placed on top of the screen, allowing for absolute charge determination[2].

Additional experimental data were taken with the Light Wave Synthesizer-20 (LWS-20) OPCPA-system then located at the Max-Planck-Institute for Quantum Optics[3]. It delivered 65 mJ on target with a pulse duration of 8 fs (FWHM). The pulses were focused to a spot diameter of 12 μm (FWHM) and guided onto a gas jet emitted by a de Laval nozzle with a 300 μm orifice.

### B. Scalings of Shock-front injection

As discussed in Sec. VI of the main text, the scaling behavior of shock-front injection was studied via experiments with different laser power on target. At 110 TW laser power on target the total accelerated charge above 80 MeV (the lower detection limit) was found to reach up to 1.2 nC. A typical electron spectrum is plotted on a linear energy scale in Figure S3.

The observed beam charge of this shock-front accelerator is much higher than in any previous experiments. In order to establish the scaling of injected charge we have varied the laser energy from 0.5 to 2 J by detuning the pump delays (cf. Fig. S4). The pulse energy for each shot is measured using a leakage diagnostic which is calibrated to the measured beam energy before the experiment. A linear fit discloses the dependence of the peak charge $Q$ on the laser power $P$ between 20 TW and 60 TW according to

$$Q_\mathrm{Peak} \approx 5.5\,\mathrm{pC} \times (P[\mathrm{TW}] - 15\,\mathrm{TW}),$$

where 15 TW is the minimum power $P_\mathrm{inj}$ for injection to occur under our conditions. As discussed in the main text, this scaling is still compatible with the results obtained at 110 TW, but overestimates the electron yield at higher peak powers. The results from PIC simulations at laser energies of up to 1 PW are shown on the right side of Fig. S4.

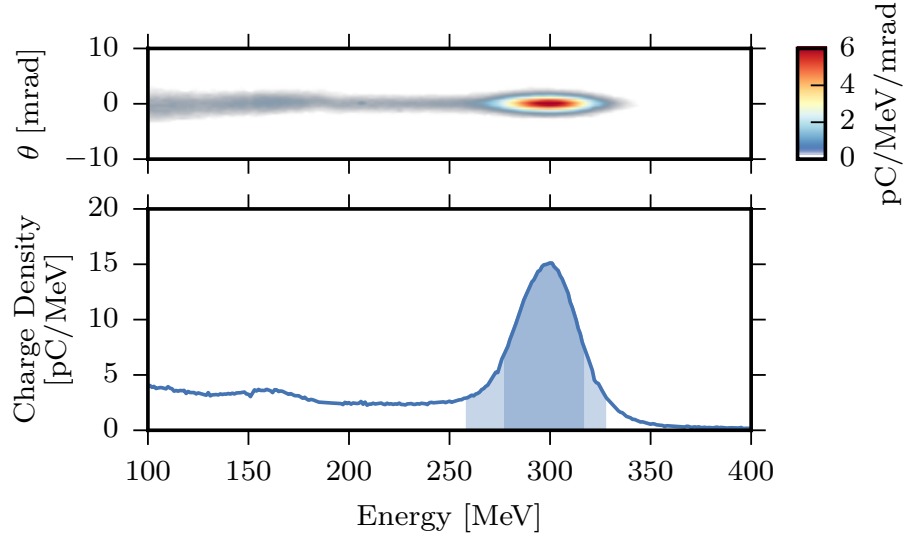

Figure S3. Electron spectrum from shock-front acceleration at 110 TW. The total accelerated charge was found to be around 1.2 nC. The peak charge around 300 MeV is beyond 600 pC. The peak charge density exceeds $15\,\mathrm{pC\,MeV^{-1}}$.

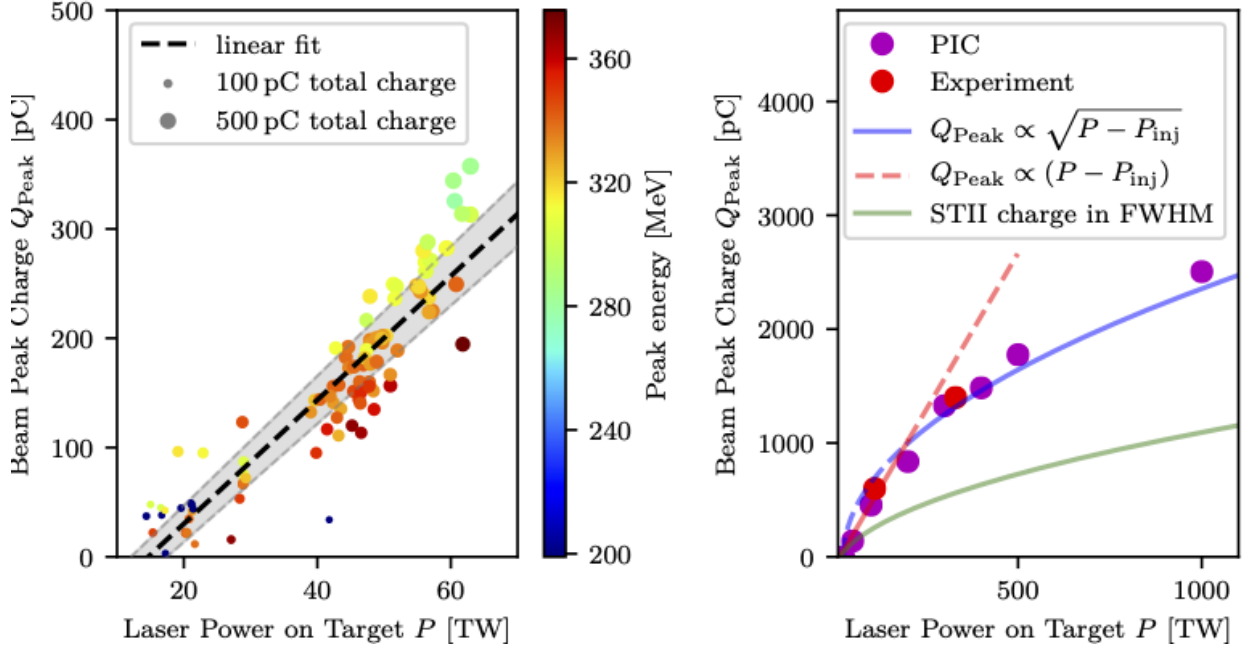

Figure S4. *Left:* Measured charges in the peak of the electron spectra vs. laser power. The size of the dots denotes the total injected charges whereas the color encodes the central energy of the peak. Note that this plot also shows beam loading effects at each individual laser power, with low electron energies for above-average charge and higher electron energies for low-charge beams. *Right:* Measured and simulated peak charges for higher laser power and corresponding scalings. The peak charge $Q_{\text{Peak}}$ in shock-front injectors is proportional to square root of the laser power $P$ for $Q_{\text{Peak}} \gtrsim 200\,\text{pC}$ and outperforms other injection mechanisms with respect to the accelerated charge like self-truncated ionization injection (STII)[4].

### C. Spectral skewness

A distinctive feature of shots with higher charge is that the beam energy spectrum is asymmetric around the peak, i.e. the energy with the highest spectral charge density. To quantify this asymmetry, we use the statistical measure of the *moment coefficient of skewness*, which is defined as the third standardized momentum

$$\gamma_3 = \text{E}\left[\left(\frac{X-\mu}{\sigma}\right)^3\right] = \frac{\mu_3}{\sigma^3} = \frac{\text{E}\left[(X-\mu)^3\right]}{(\text{E}\left[(X-\mu)^2\right])^{3/2}} = \frac{\kappa_3}{\kappa_2^{3/2}}, \tag{S1}$$

with $X$ being the respective random variable (the energy spectrum in our case), $\mu$ being the standard deviation, E is the expectation operator, $\mu_3$ is the third central momentum, $\sigma$ the standard deviation and $\kappa_i$ denotes the i-th cumulant. The skewness coefficient $\gamma_3 = 0$ for distributions which are symmetric around the average energy $\mu$. If the electron spectra have a peak at high and a tail at lower energies $\gamma_3$ becomes negative, whereas spectra with $\gamma_3 > 0$ have their peak at the low end and a tail at higher energies, cf. Fig. S5. An energy spectrum with negative skewness is the typical distribution one would expect from an initially negatively chirped electron bunch (generated for example via continuous self-injection of electrons) entering dephasing. Instead, beam loading should lead to a strong peak at low energies and a tail beyond the mean energy for an initially positively chirped electron phase space (generated for example via shock-front injection) which results in a positively skewed spectrum as shown later in Fig. S9. For the ATLAS-300 data, we find an almost linear correlation with a Pearson correlation coefficient of 0.63 between skewness ($\gamma_3$) and peak charge according to

$$\gamma_3 \approx (2.7 \pm 0.4) \times Q[\mathrm{nC}] - (0.18 \pm 0.06) \,, \tag{S2}$$

where $Q$ denotes the peak charge contained in the spectrum given in nC. Fig. S6 shows the calculated skewness for each shot of the LWS-20 data in Fig. 3(a) (*left*) and ATLAS-300 data in Fig. 3(c) (*right*). The predominantly positively skewed spectra and the positive correlation of $\gamma_3$ with the peak charge support our understanding of beam loading in shock-front injectors (cf. main part Sec. III).

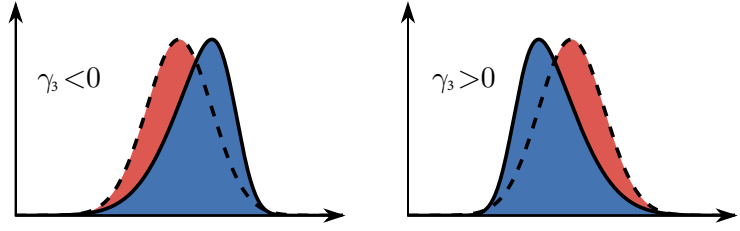

Figure S5. Illustration of skewed spectra. Negatively skewed (left, blue) and positively skewed spectrum (right, blue) compared to a normal (Gaussian) distributed spectrum (red).

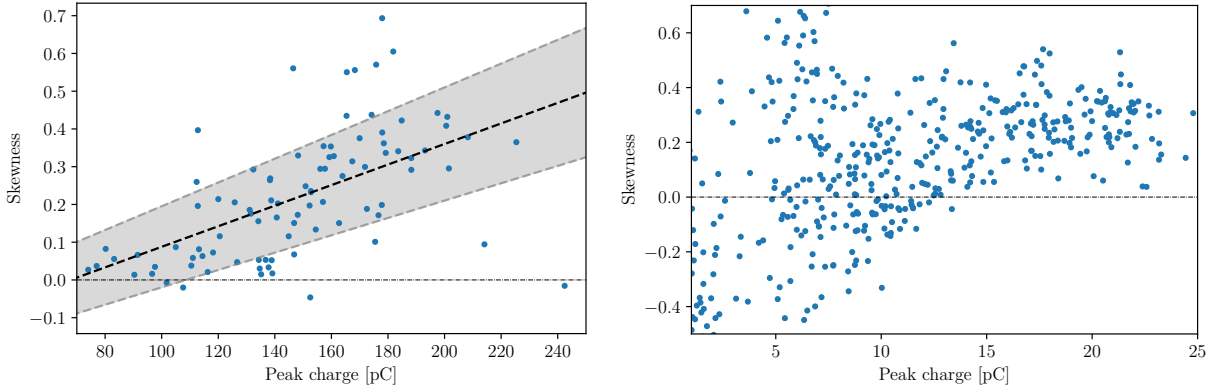

Figure S6. *Left:* Calculated skewness for each individual spectrum of the ATLAS-300 data set depicted in Fig. 3(c) in the main part. Almost all spectra are positively skewed which clearly hints at beam loading effects. *Right:* Calculated skewness for each individual spectrum versus peak charge of the LWS-20 data set depicted in Fig. 3(a) in the main part. Again, all spectra at higher charges are positively skewed. However, due to the poor signal-to-noise ratio, we cannot fit a linear trend line for low peak charges in this case.

## II. SIMULATION STUDIES

All simulations for the paper were performed using the quasi-3D PIC code FBPIC[5], with some using a boosted-frame technique with Galilean algorithm[6,7]. An overview of the simulations and their parameters done in the scope of this work is given in Table S1.

### A. Parameter scan for different regimes of laser and beam-driven wakefields

To give the reader a qualitative understanding of the transition between wakefields dominated by either the laser or the beam driver, we have performed a 2D scan of electron beam charge and laser intensity, shown as density snapshots in Fig. 1. Rows $1-6$ use the same laser power ($100\,\text{TW}$) at a duration of $\Delta t = 30\,\text{fs}$, while the laser waist is varied from $w_0 = 12\,\mu\text{m}$, corresponding to $a_0 = 4.0$, to $w_0 = 48\,\mu\text{m}$, i.e. $a_0 = 1.0$. The waists for the respective rows are $[12, 15, 18, 24, 36, 48, \infty]\,\mu\text{m}$ (top to bottom). Note that the laser is absent in the last row, corresponding to a purely beam driven case. The beam charge along the columns increases as $[0, 100, 300, 500]\,\text{pC}$ (left to right). The background plasma density is $n_e = 3 \times 10^{18}\,\text{cm}^{-3}$ in all cases.

As discussed in the main text, the results from Fig. 1 are not only relevant for the specific parameters given above. Instead, they illustrate the general interplay between intense laser drivers and electron drivers.

| Fig. | Code | $\Delta z$ [nm] | $\Delta r$ [nm] | Boost |
|------|------|-----------------|-----------------|-------|
| 1    | FBPIC-0.15 | 32 | 167 | no |
| 3    | FBPIC-0.9  | 28 | 160 | no |
| 4    | FBPIC-0.15 | 40 | 400 | no |
| 5    | FBPIC-0.15 | 40 | 400 | $\gamma = 5$ |
| S7   | FBPIC-0.15 | 32 | 266 | $\gamma = 5$ |
| S8   | FBPIC-0.9  | 28 | 160 | no |
| S9   | FBPIC-0.9  | 28 | 160 | no |
| S10  | FBPIC-0.15 | 40 | 400 | $\gamma = 5$ |
| S11  | FBPIC-0.15 | 40 | 400 | $\gamma = 5$ |

Table S1. Summary of the simulation parameters for each figure.

### B. Simulating shock-front injection

We have performed an extensive study of shock-front injection. The simulation windows had a typical length of 70 µm with 80 µm radius. The number of grid points in each direction varied such that we had resolutions between 28 nm and 40 nm in longitudinal direction and 160 nm to 400 nm in the radial dimension (cf. Table S1). The laser pulse duration was chosen to be 30 fs with a spot size of $w_0 = 17$ µm. The plasma density profile was adopted according to the interferometric measurements (cf. Fig. S2). As it is difficult to resolve the width of the shock, we have chosen a transition length of $\lambda_p$.

### C. Adaptive macro-particle sampling

For accurate beam loading simulations it is essential to get a good estimate of the injected charge, which requires a high density of macro particles in the simulations. However, this usually leads to very long simulation times and high memory load, prohibiting extensive or high-resolution parameter scans. To circumvent this issue, we take advantage of the fact that electron injection in our accelerator is localized around the density transition (cf. Fig. S8). This allows us to locally increase the macro particle resolution to get higher statistics and thus converging beam parameters. For the simulations shown in the main text, we used a density of 64 macro particles per cell (ppc) in the region around the shock. In contrast, the outer regions have been sampled with lower macro particle density (typically $4 - 8$ ppc). As particles from these regions do not contribute to the total

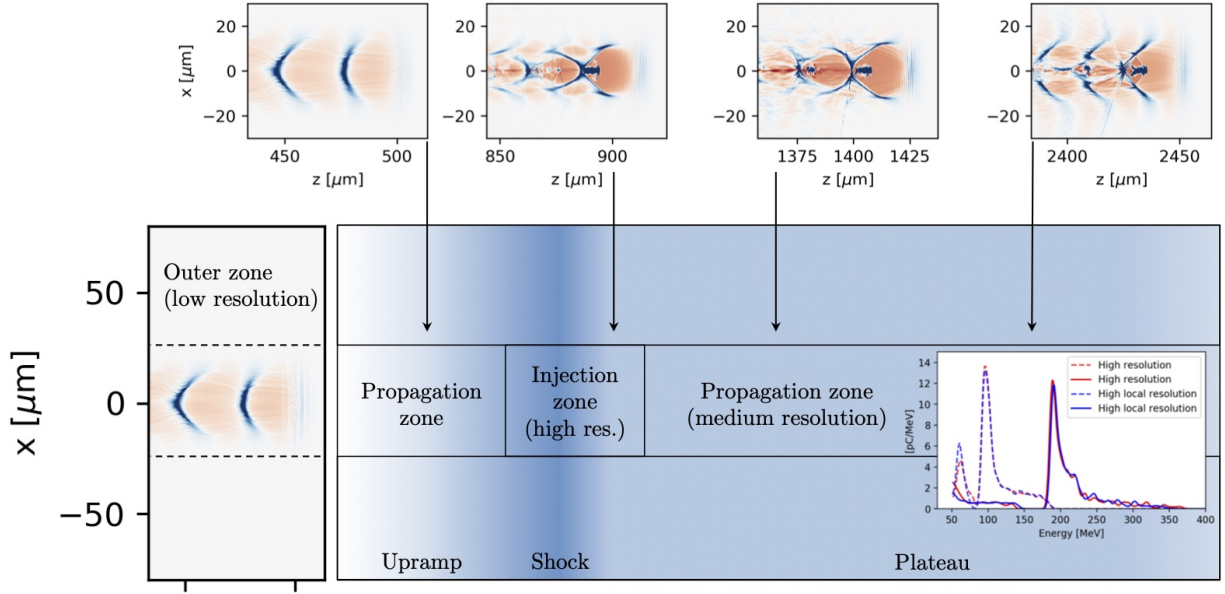

Figure S7. The whole PIC simulations are composed of several regions which are sampled with different macro particle density. The most relevant zone - the area around the shock where injection occurs - is sampled with highest resolution. The inset on the right compares a high resolution simulation with constant high macro particle density everywhere (red) to a simulation where only the injection zone is accurately sampled (blue). Their electron spectra differ only marginally as exemplarily depicted for two different snapshots (dashed and solid lines).

accelerated charge, the resolution here can be reduced for an overall shorter computation time.

This technique is also suitable in combination with boosted-frame simulations, where the reduced sampling resolution along the axis of the boost can easily be compensated for by increasing the number of particles initiated along the same axis. This allows for simulation times of the order of $1-2$ hours (quasi-3D) or 15 minutes (quasi-3D and boosted frame) for one 3-mm-laser-plasma simulation on a single TESLA V100 GPU. It should be noted that this technique is less suitable for multi-CPU or multi-GPU simulations as it leads to load-balancing issues.

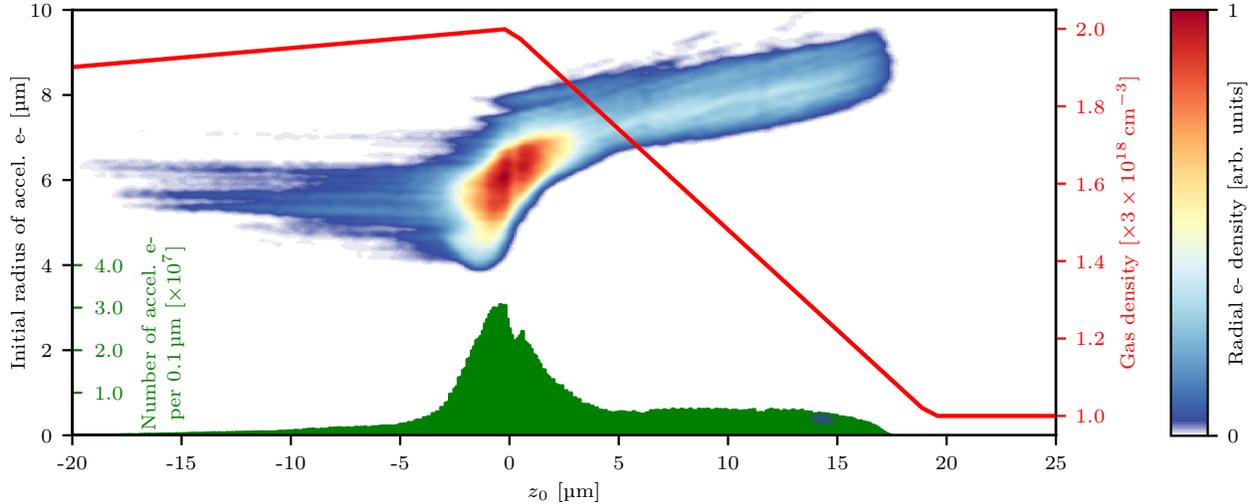

Figure S8. Plot depicting the initial position of captured electrons. The green histogram shows that only electrons initially located less than one plasma wavelength before the shock and during the density downramp (red line) are captured and accelerated. Therefore it is sufficient to only finely sample the region around the shock and the downramp to get a decent simulated charge estimate. The initial radial distance of the accelerated electrons is less than $w_0$. In these simulations we chose $w_0 = 17\,\mu\text{m}$.

### D. Injection region for shot front injection

To get a decent estimate for the extension of the accurately sampled 'injection zone' in Fig. S7 and to gain better insight into the shock-injection process itself, we have tagged the macro particles and determined the initial position of the electrons that end up being accelerated. The result of the analysis is shown in Fig. S8 and clearly shows that most of the injected electrons indeed originate from the beginning of the downramp, but there are also electrons being injected which are located less than a plasma wavelength away from the start of the downramp. As expected from the formation of the wakefield due to the transverse ponderomotive potential[1], most electrons that are injected are initially located off axis.

### E. Focus scan

Different parameter scans were performed including a variation of the longitudinal focal spot position from $-1\,\text{mm}$ to $1\,\text{mm}$ relative to the shock front in 21 equidistant steps. The position of the focal spot changes the amount of accelerated charge as depicted in Fig. S9. As discussed in the main text, this change not only takes place on the scale of the Rayleigh length ($\sim 1.1\,\text{mm}$), but even simulations with a focus shift of only $0.2\,\text{mm}$ exhibit a significant difference in charge. Note that this is even more pronounced the further we are away from optimal injection ($z \approx -0.1\,\text{mm}$).

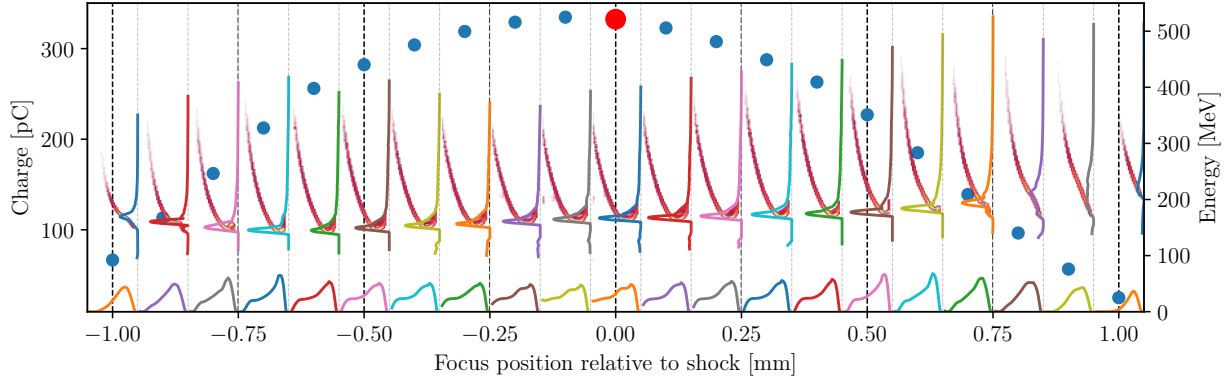

Figure S9. Simulation study of the influence of the longitudinal focal position relative to the shock front. The corresponding Rayleigh length is $L_R = 1.1\,\text{mm}$. The beam charge at each position is plotted as blue dot. The red dot marks the position $L_\text{foc} = L_\text{shock}$ and was used as reference case for all other scans. At each position, the longitudinal phase space as well as the current profile (bottom) and integrated spectrum (right) are given.

Due to the charge dependence of beam loading, the phase space distribution and final electron energy spectra also change significantly. Notably, spectra with high charge show a strong nonlinear energy chirp and a positively skewed energy spectrum. In turn, low charge cases (e.g. focus at $z_\text{foc} \approx 0.8\,\text{mm}$ have a quasi-linear chirp and a more symmetric spectrum.

Note also that the energy spectra clearly follow the shape of the phase space (i.e. the energy peak is always situated at the 'turning point' of the phase space) and thus, seem much less sensitive to the bunch current (shown at the bottom).

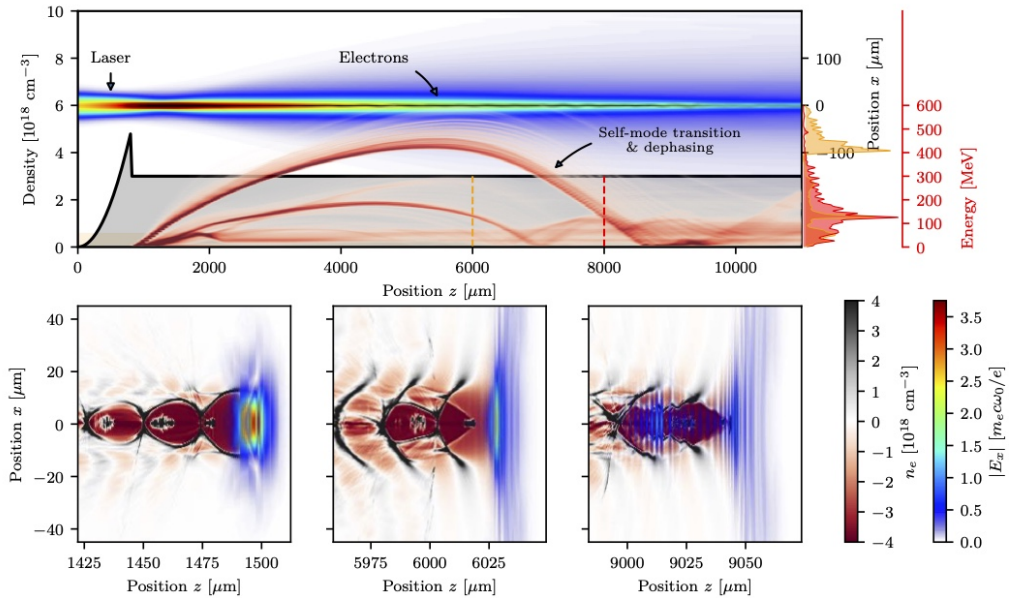

Figure S10. Propagation of a laser and two shock-injected bunches in a long plasma with constant background density. Analogously to Fig. 5 of the main text, the top frame summarizes the evolution of the laser intensity and the electron beam spectrum, while the lower frames show snapshots of the plasma density and the laser field at the injection, electron dephasing and mode transition locations. The simulations show how the bunches enter the dephasing region of the laser wakefield while also transitioning to a beam-dominated wakefield. The spectra for different acceleration length (here shown for $z = 6\,\mathrm{mm}$ and $z = 8\,\mathrm{mm}$) are therefore difficult to interpret. Moreover, in contrast to the case of a controlled transition that show the formation of a clear bubble-structure behind the electron bunch (see Fig. 5 and Fig. S11), the wakefield structure is still strongly influenced by the residual laser pulse in this case.

### F. Start-to-end simulations for self-mode transition and beam-dominated regime

For these start-to-end simulations we again employed the FBPIC code, but due to the considerable larger propagation distances of $z_{\mathrm{max}} = 11\,\mathrm{mm}$ we switched to a boosted-frame simulation with $\gamma_{\mathrm{boost}} = 5$. Furthermore, we use a larger radial size for the simulation window ($r_{\mathrm{max}} = 200\,\mathrm{\mu m}$), while the longitudinal length of the window is $90\,\mathrm{\mu m}$. The entire simulation window is resolved by $n_z \times n_r = 2250 \times 500$ grid points. Regarding the macro particle density, we use a resolution of $\mathrm{ppc}_z \times \mathrm{ppc}_r \times \mathrm{ppc}_\theta = 4 \times 2 \times 12$ in the inner part ($r < 50\,\mathrm{\mu m}$), with $\mathrm{ppc}_z \times \mathrm{ppc}_r \times \mathrm{ppc}_\theta = 16 \times 8 \times 24$ around the shock. The outer regions of the simulation are sampled with $\mathrm{ppc}_z \times \mathrm{ppc}_r \times \mathrm{ppc}_\theta = 1 \times 1 \times 6$, as is the ion species. Furthermore we use the modes $m = 0 - 3$ for Fourier decomposition along the poloidal direction. To avoid reflections of the laser at the boundary we set absorbing boundaries in transverse and longitudinal direction.

The simulated density profile is shown in the respective simulation figures, i.e. Fig. S10, Fig. S11 and Fig. 5.

Fig. S10 shows a simulation for the self-mode transition regime, i.e. a single long plasma stage. In this situation the mode transition is superimposed with dephasing, leading to an ambiguous signature in the energy spectrum. Furthermore, even after 9 mm of propagation in the plasma the laser still affects the wakefield formation significantly, resulting in a more or less diamond-shaped wake instead of the typical bubble.

This problem is avoided in the dual-jet case, where laser and beam evolution are decoupled. Here Fig. S11 and Fig. 5 show simulations with different total charge in the driver (due to different focusing positions), which results in differently pronounced acceleration of the witness. Note that we initially used perfect vacuum in the transition zone between both jets ($z \approx 4-8$ mm). However, we noticed that the electron beam divergence was much larger than in the experiment in this case and thus, only a small fraction of the second beam was caught as witness in the second stage. This clearly did not accurately represent the experimental data, where we measure a divergence of $\sim 1$ mrad. In the simulations, we noticed a strong dependence of the beam divergence with the shape of the downramp. We therefore adapted the downramp of the jet to match the divergence of the experiment.

14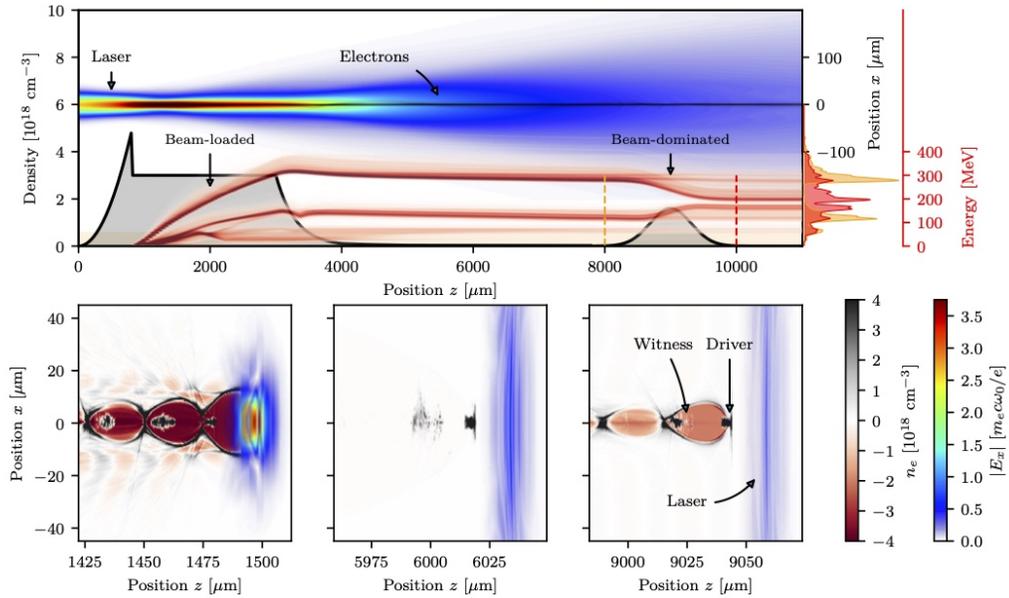

Figure S11. Start-to-end simulation showing the laser and beam evolution while propagating through two subsequent gas jets. The total simulation covers a distance of 11 mm. The simulation shown here differs from Fig. 5 in the higher charge of the first electron beam (here 125 pC) resulting in a higher energy transfer from the driver to the witness bunch in the second gas jet. Please also note the different form of the final electron spectrum compared to this figure, i.e. that the beam spectra start to overlap. In experiment, such a case is therefore difficult to analyse.

[1] H. Ding, A. Döpp, M. Gilljohann, J. Götzfried, S. Schindler, L. Wildgruber, G. Cheung, S. M. Hooker, and S. Karsch, Nonlinear plasma wavelength scalings in a laser wakefield accelerator, Physical Review E **101**, 023209 (2020).

[2] T. Kurz, J. P. Couperus, J. M. Krämer, H. Ding, S. Kuschel, A. Köhler, O. Zarini, D. Hollatz, D. Schinkel, R. D'Arcy, J.-P. Schwinkendorf, J. Osterhoff, A. Irman, U. Schramm, and S. Karsch, Calibration and cross-laboratory implementation of scintillating screens for electron bunch charge determination, Review of Scientific Instruments **89**, 093303 (2018).

[3] D. Herrmann, L. Veisz, R. Tautz, F. Tavella, K. Schmid, V. Pervak, and F. Krausz, Generation of sub-three-cycle, 16 TW light pulses by using noncollinear optical parametric chirped-pulse amplification, Optics Letters **34**, 2459 (2009).

[4] J. P. Couperus, R. Pausch, A. Köhler, O. Zarini, J. M. Krämer, M. Garten, A. Huebl, R. Gebhardt, U. Helbig, S. Bock, K. Zeil, A. Debus, M. Bussmann, U. Schramm, and A. Irman, Demonstration of a beam loaded nanocoulomb-class laser wakefield accelerator, Nature Communications , 1 (2017).